# Synthesis and different Optical properties of $Gd_2O_3$ doped sodium zinc tellurite glasses

Buddhadev Samanta

Department of Physics & Astronomy, National Institute of Technology

Rourkela, Odisha, 769008.

515ph6010@nitrkl.ac.in

### Abstract

Optical properties of a series of $Gd_2O_3$ doped sodium zinc tellurite [$xGd_2O_3$-(0.8-x) $TeO_2$-0.1$Na_2O$-0.1ZnO] glasses prepared by conventional melt quenching method have been studied. UV-Vis spectrophotometer was used in the integrating sphere mode at room temperature to study the effect of $Gd_2O_3$ doping on the optical band gap, refractive index, dielectric constant and susceptibility within the wavelength range from 230-800 nm. Density is measured using Archimedes principle for the calculation of molar volume, molar refraction, polarizability and metallization criterion of the glasses.

## 1. Introduction

Tellurite glasses are very important because of their technological and scientific applications [1]. Now a days, rare-earth doped transparent tellurite glasses and glass ceramics are of increasing interests in various optical applications [2] because of their superior optical properties, such as high refractive index, high dielectric constant, chemical durability, electrical conductivity, wide band infrared transmittance and large third order non-linear susceptibility [2]. $TeO_2$ is a conditional glass former [3-5] and hence requires small molar percentage addition of oxides and halides to form glasses. It also has the lowest phonon energy [2] among the oxide glass formers, as well as good rare earth ion solubility, which makes it efficient host for dopant ions like rare earths, allowing a better environment for radiative transitions. The dopant ions maintain majority of their individual properties in the glass matrix. Due to low melting point ($T_m$) and glass transition temperature ($T_g$) [6, 7], preparation of tellurite glasses are comparatively easier than other glass families. Optical absorption in solids occurs due to the absorption of photon energy by either the lattice (or phonon) or by electrons. The phonon absorption will give information about atomic vibrations involved and this absorption of radiation normally occurs in the infrared region of the spectrum.

Optical absorption is a useful method for investigating optically induced transitions and to get information about the band structure and energy gap of non-crystalline materials. The principle of this technique is that a photon with energy greater than the band gap energy will be absorbed [8]. The optical band gap is determined from absorbance (α) data as a function of Energy (hv) by using the relation [9, 10] given by

$\alpha hv = A\ [hv-E_g]^{\ r}$ (1)

where 'A' is a constant, '$E_g$' is the optical band gap of the material and the exponent 'r' depends on the type of transition and have values 1/2, 2, 3/2 and 3 corresponding to the allowed direct, allowed indirect, forbidden direct and forbidden indirect transitions respectively. Refractive index (n), dielectric constant ($\epsilon_r$) and susceptibility (χ) are also the important properties for optical glasses and a large number of researchers have carried out investigations about the relation of refractive index, dielectric constant, and susceptibility separately with glass composition. It is generally recognized that the refractive index, dielectric constant, susceptibility of many common glasses can be varied by changing the base glass composition [2].

In this article we have studied the change in optical band gap, refractive index, dielectric constant, susceptibility, density, molar volume, molar refraction, metallization criterion and polarizability due to increase of $Gd_2O_3$ content in the glass network.

## 2. Materials and methods

Mixture of 10 gm batch with proper stoichiometric mixing of $TeO_2$, $Na_2CO_3$, ZnO and $Gd_2O_3$ powders, procured from Sigma Aldrich has been taken for the preparation of glass samples. Table 1, below shows the name assigned to the glass samples and the stoichiometry at which they were mixed.

### Table 1

Name and composition of different samples

| **Sample name** | **Mole% of $Teo_2$** | **Mole% of $Na_2O$** | **Mole% of ZnO** | **Mole% of $Gd_2O_3$** |
|---|---|---|---|---|
| NZTGO | 80 | 10 | 10 | 0 |
| NZTG1 | 79 | 10 | 10 | 1 |
| NZTG2 | 78 | 10 | 10 | 2 |
| NZTG3 | 77 | 10 | 10 | 3 |

### Glass preparation

$Gd_2O_3$ doped [$xGd_2O_3$ - (0.80-x) TeO2 - $0.10Na_2O$ -0.10 ZnO] with x=0.00, 0.01, 0.02, 0.03 glass samples were prepared by conventional melt quenching technique. First of all each glass sample was taken in an alumina crucible and melted at ~ 890°C inside an electrical furnace [CBC Power System] for 2 hour.Then it was poured onto a stainless still plate and quickly quenched in air. After quenching, the glass was annealed at 250°C for about 3 hours and then normally cooled down to room temperature to release the thermal and mechanical stress [11-13] from the samples.

## Instrument details:

Ultra violate-visible spectroscopic study was done using UV-VIS-3092 manufactured by Lab India (Serial No- 20-1950-01-0025). Resolution of spectrophotometer is 1.75 (for Toluene and Hexane). Deuterium lamp is used as source (656.1 nm & 486.0 nm) producing wavelength range from 230 to 850 nm with bandwidth 2.0 nm. Integrating sphere (Model 1S19-1) is used for execution of diffuse reflectance data in which Incident angle of Sample is $0^0$ and incident angle of reference is $8^0$. Slit width is 5 nm. Minimum dimension of integrating sphere for diffuse reflection is [width-15mm, height 25mm]. Diameter of the sphere is 58 mm. Barium-Sulphate [$BaSO_4$] is used as a background substrate. Photomultiplier is used as a detector in this UV-VIS spectrophotometer.

## Band gap measurement in diffuse reflection method by using 'Kubelka-Munk Theory'

At first each sample was being powdered carefully using a morter and then UV-Vis spectrophotometer (Lab India UV-3090), in the integrating sphere mode was used over a wavelength range of 230-850 nm in order to measure the diffuse reflectance of the as prepared glasses. 'Kubelka-Munk Theory' [14] has been used to calculate the optical band gap for these glasses from the diffused reflectance data.

## The Kubelka–Munk Theory:

Let us consider a plane parallel layer of thickness 'L', where both scattering and absorption of radiation are possible. Monochromatic radiation flux 'I' in the –x direction is incident on such layer. The layer can be considered as sum of infinitesimal sub-layers of thickness 'dx' (shown in Figure 1)

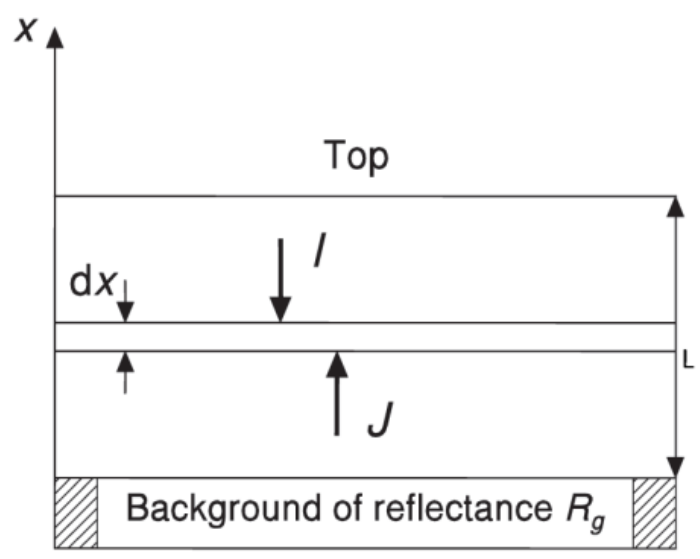

**Figure1**

The diffuse radiation flux can be represented in the direction of +x and –x as '**J**' and '**I**' respectively. At the time of passing through layer 'dx' the downward flux decreased by amount IKdx due to absorption and increased by amount ISdx due to scattering. Similarly for outward flux 'J' we can think of that. So differential equation can be represented for radiation flux

$$-dI = -(K+S)\,Idx + JSdx \quad (2)$$

$$dJ = -(K+S)\,Jdx + ISdx \quad (3)$$

Here Kand Sare the absorption and scattering coefficient of the sample respectively. Introducing a new term R=J/I, reflectance of the sample.Byregorous calculationit has been found that

$$dR / [R^2 - 2R(K/S + 1) + 1] = Sdx \quad (4)$$

Now integrating with in the limit from(x=0, $R=R_0$) to (x=L, $R=R_\infty$)

$$\int_{R_0}^{R_\infty} \frac{dR}{(R-a)^2-b^2} = SL \quad (5)$$

Here a= 1+K/S and b= $(a^2-1)^{0.5}$. Result of the integration gives

$$R_\infty = \frac{1-R_0[a-b\mathrm{Coth}(bSL)]}{[a-R_0+b\,\mathrm{Coth}\,(bSL)]} \quad (6)$$

Here $R_\infty$ is the reflectance of the layer over a background of reflectance $R_0$.Themeasurements are in such way that further increase in thickness will fail to change the reflectance. For black background $R_0 = 0$ and for large enough thickness L must be tending to infinite. So Coth (bSL) →1.From equation no (5) it has been found that

$$R_\infty = 1/(a+b) = \frac{S}{K+S+\sqrt{K(K+2S)}} \quad (7)$$

So last of all we find

$$K/S = (1-R_\infty)^2 / 2R_\infty = F(R_\infty) \quad (8)$$

Here F ($R_\infty$) is remission or Kubelka–Munk (K–M) function. Instead of writing $R_\infty$, R is suitable here.

So $$F(R) = (1-R)^2 / 2R \quad (9)$$

K/S = $(1-R_\infty)^2 / 2R_\infty$ = F ($R_\infty$), Where K and S are absorption and scattering coefficient respectively. If we assume the scattering from the material remains constant for the entire wavelength range, Kubelka-Munk function contributes the absorption co-efficient. So the 'Kubelka-Munk function' is proportional to the absorption co-efficient (α) [14], [F (R) ∝ α].UV-Vis spectra after been processed with Kubelka-Munk function for the sample NZTG1 is shown in Figure 2.

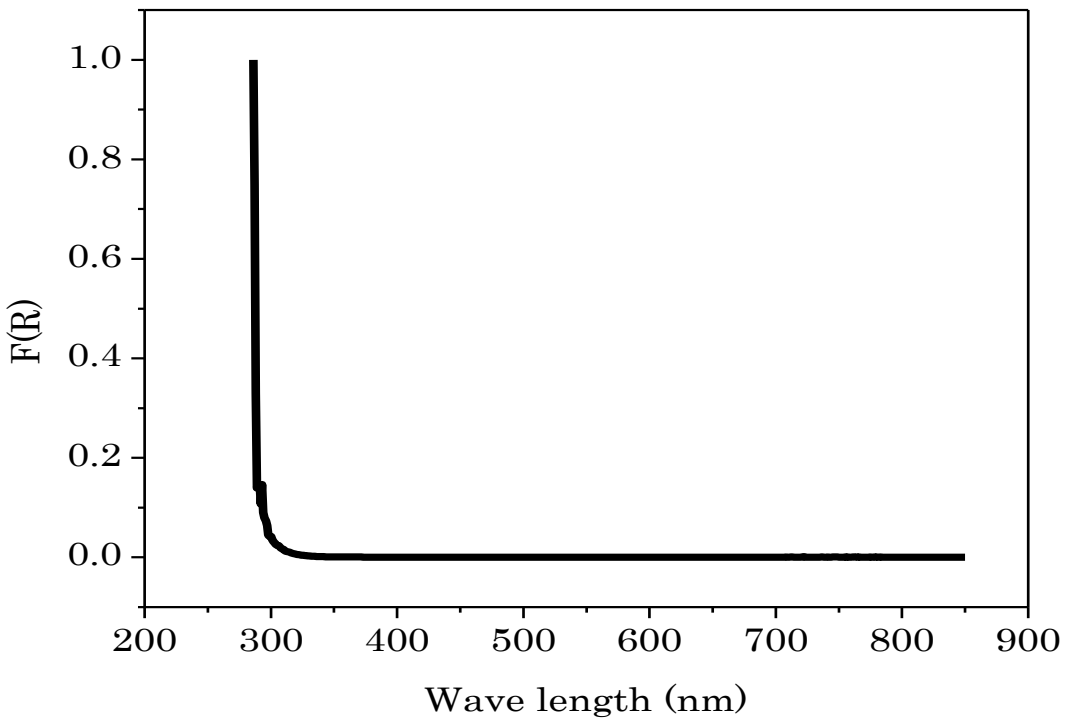

**Figure 2.** Calculated Kubelka-Munk function F(R) of sample NZTG1 as function of wavelength

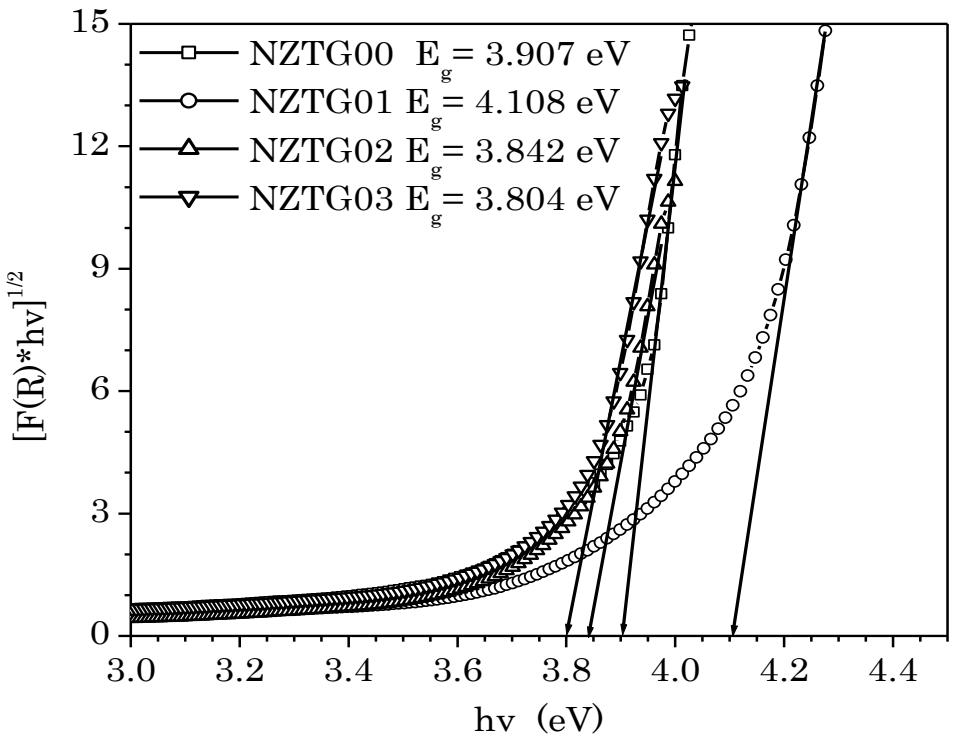

**Figure3.** Estimating the optical band gap from reflectance spectra processed byKubelka-Munk function for different glass compositions.

## Band gap measurement:

As the ‘Kubelka-Munk function’ is proportional to the absorption co-efficient (α) Equation no (1) suggests that

$$[F(R)*hv]^{1/r} \propto [hv - E_g] \qquad (10)$$

By taking the value of r = 2, $[F(R)*hv]^{1/2}$vs. hv plot for the allowed indirect band gap measurement is shown in the Figure 3. The point of intersection of the tangent to linear vertical region and the horizontal axis (hv-axis) gives the band gap value. Figure 4 shows the variation in the optical band gap of the glasses with the variation of mole fraction in the composition of glasses.

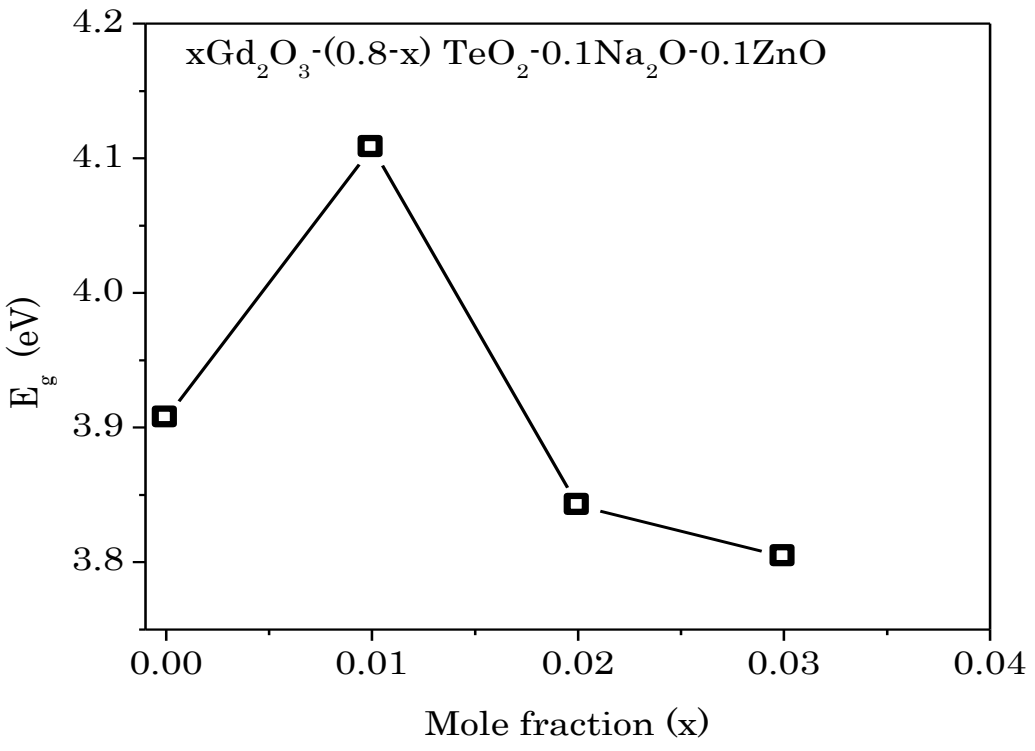

**Figure4**. Influence of molar fraction of $Gd_2O_3$ on band gap of investigated glasses

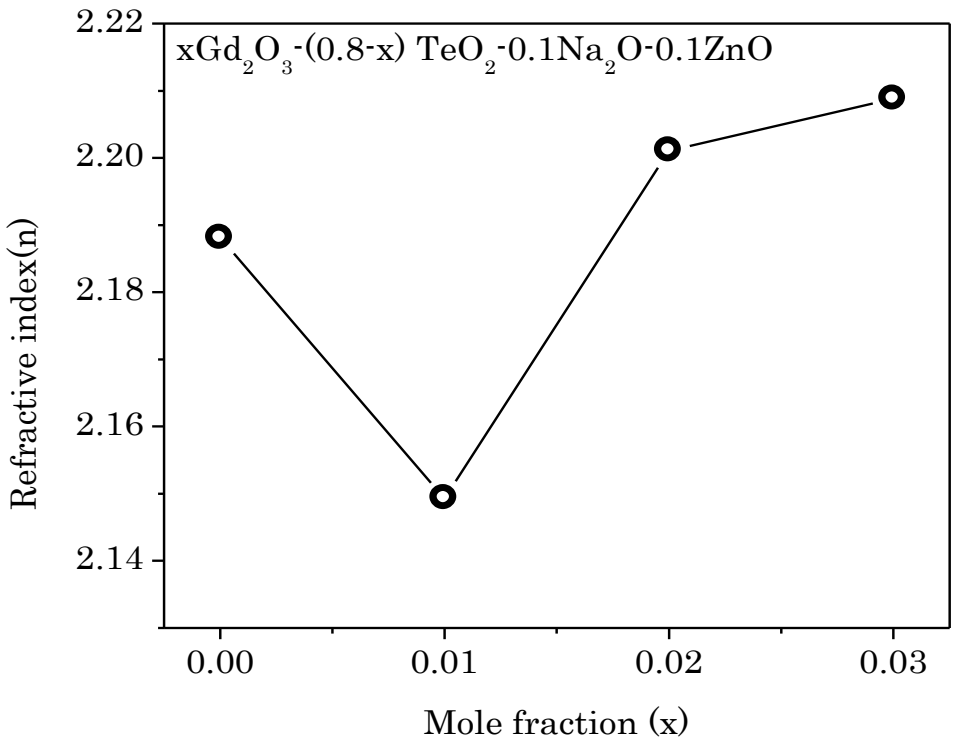

**Figure5.**Influence of molar fraction of $Gd_2O_3$ on refractive index of investigated glasses

## Refractive index measurement

Experimentally measured optical band gap and the refractive index (n) are related by a relation given by Dimitrov and Sakka (1996) [15-19], which is given by

$$(n^2-1)/ (n^2+2) =1-(E_g/20)^{1/2} \qquad (11)$$

Figure 5 shows the variation of linear refractive index (n) with changing molar fraction of $Gd_2O_3$

## Linear dielectric constant and susceptibility measurement

Static or linear dielectric constant can be measured using the relation [20, 21] given by

$$\epsilon_r = n^2 \qquad (12)$$

Linear susceptibility (χ) is related with dielectric constant as

$$\epsilon_r = 1+\chi \qquad (13)$$

Figure 6 shows the variation of linear dielectric constant and susceptibility with changing molar fraction of $Gd_2O_3$.

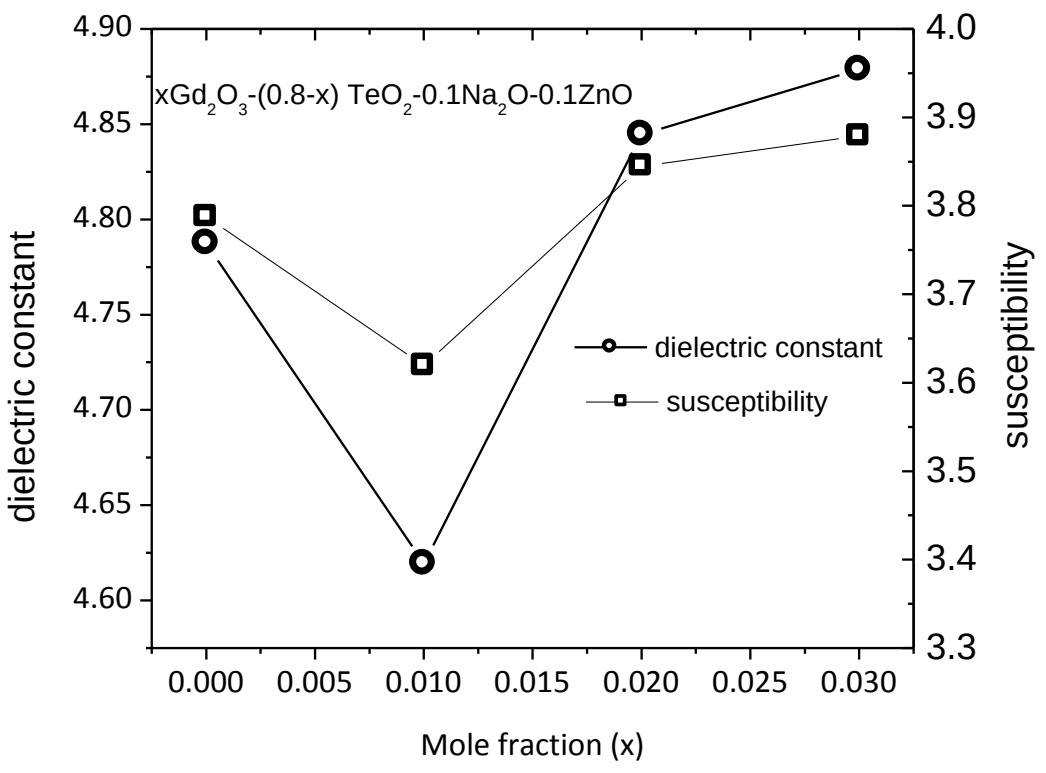

**Figure 6.** Influence of molar fraction of $Gd_2O_3$ on dielectric constant and susceptibility of investigated glasses.

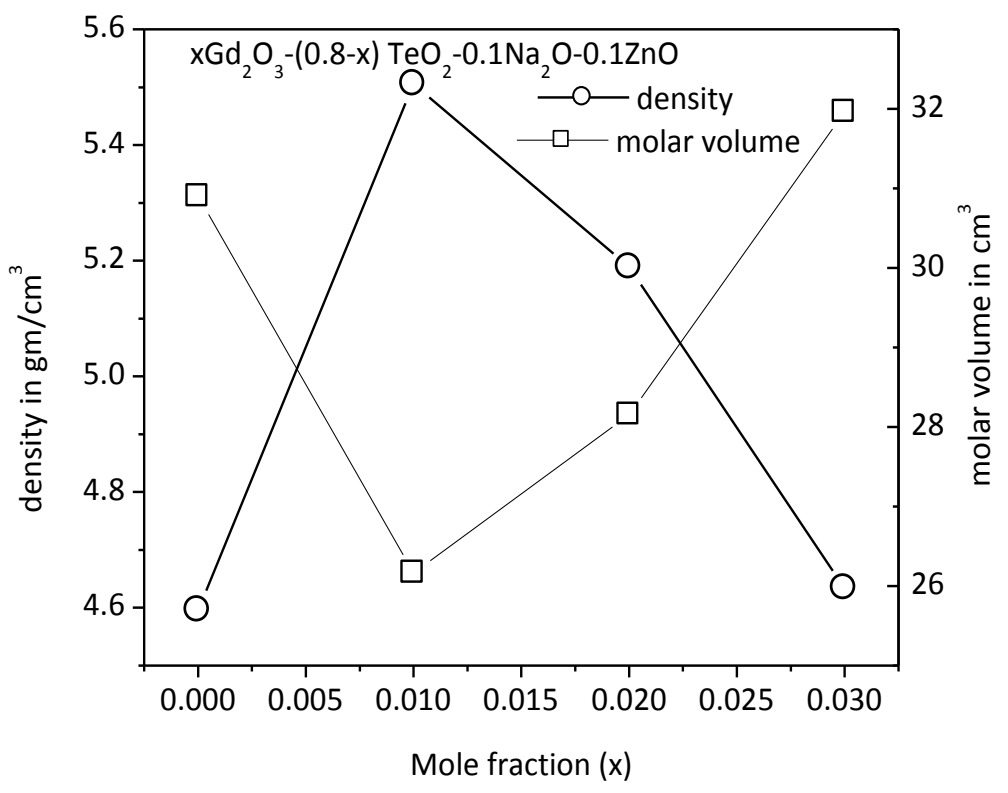

**Figure 7** Influence of molar fraction of $Gd_2O_3$ on density and molarvolume of investigated glasses

## Calculation of molar refraction ($R_M$), molar volume ($V_m$) and polarizability ($\alpha_e$) through measurement of density (ρ)

The equation which relates the polarizability ($\alpha_e$) and refractive index (n) is given by [8]

$$[(n^2-1)/ (n^2+2)]\ V_m = (4/3)N\alpha_e \qquad (14)$$

So $\alpha_e = 3*[(n^2-1)/ (n^2+2)] *V_m/ (4\pi N)$ (15)

Where '$V_m$', 'N', '$\alpha_e$' are molar volume, Avogadro number and the polarizability respectively. When equation no (15) is expressed in terms of the specific mass or density ρ, it reduces to

$$[(n^2-1)/ (n^2+2)]\ M/\rho = R \qquad (16)$$

The above equation describes the specific refraction, R, of the material. When, M is the molar weight of the material and M/ρ the molar volume ($V_m$) the above equation becomes

$$[(n^2-1)/ (n^2+2)]\ M/\rho = R_M \text{ (the molar refraction)} \qquad (17)$$

Equation (11) and (15) combined to give the value of polarizability using band-gap, molar volume and Avogadro number such that

$$\alpha_e = 3*[1-(E_g/20)^{1/2}]*V_m/(4\pi N) \quad (18)$$

Molar volume can be calculated by calculating the molar weight (M) and density (ρ) of the glass samples as $V_m = M/\rho$ (19)

Density is measured using Archimedes principle using acetone as an immersion liquid (density of acetone is $0.791 gm/cm^3$ at room temperature). Figure 7 shows the influence of changing mole fraction of $Gd_2O_3$ on the density and molar volume. Figure 8 shows the variation of molar refraction and polarizability with changing mole fraction of $Gd_2O_3$ in the glass composition. Figure 9 shows the variation of optical band gap with density of glass material.

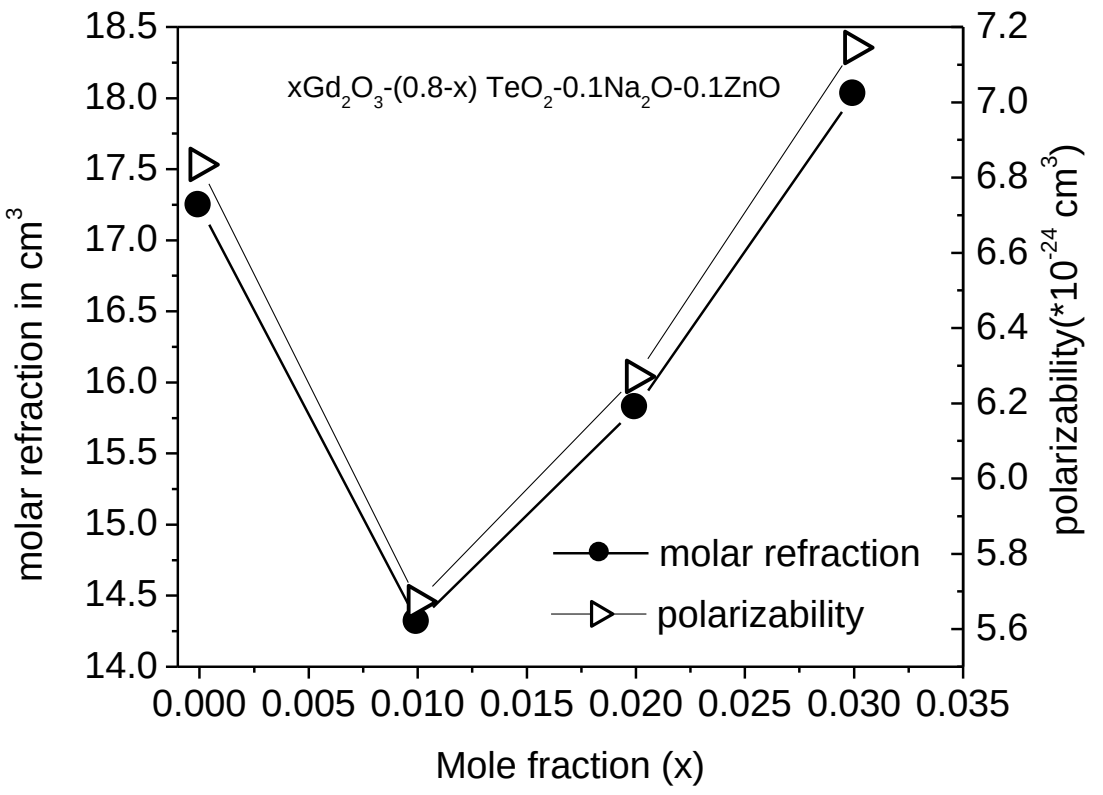

**Figure 8.** Influence of molar fraction of $Gd_2O_3$ on molar refraction and polarizability of investigated glasses

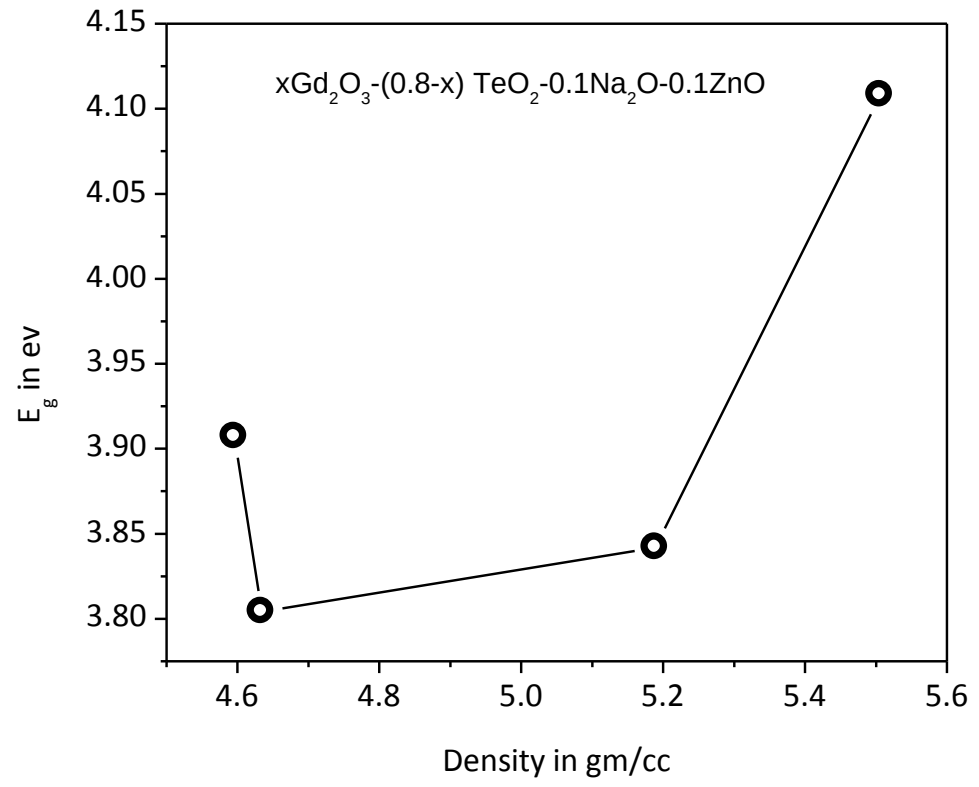

**Figure 9.** Infuence of density on Optical band

## Metallization criterion

Metallization criterion was explained by Dimitrov and Komatsu using Lorentz-Lorenz equation on the basis of optical band-gap and static refractive index given by [24]

$$M' = 1-(n^2-1)/(n^2+2) = (E_g/20)^{1/2} = 1-(R_M/V_m) \quad (19)$$

Transition to metal state occurs when the above equation becomes zero. Metallic and non-metallic nature of oxide glasses can be predicted by the relation: M'>1(for metallic) and M'<1 (for non-metallic) [24]. In Table 2 we have shown the variation of metallization criterion with increase mole fraction of $Gd_2O_3$.

## Calculation of Number density of rare-earth ions (N), Polaron radius ($r_p$), Inter ionic distance ($r_i$).

The number density of rare-earth ions can be determined using the formula given by [25]

$$N = [6.023\times10^{23}\times\text{mol\% of cation}\times\text{valency of cation}]/V_m. \quad (20)$$

The obtained values of N are used to calculate the polaron radius ($r_p$) and ionic radius ($r_i$) using the relations [25, 26],

Polaron radius $r_p=0.5(\pi/6N)^{1/3}$ (21)

Inter ionic distance $r_i = (1/N)^{1/3}$ (22)

In table 2 the variations of N, $r_p$ and $r_i$ with changing $Gd_2O_3$ moler fraction have been shown. Figure 10 shows the variation of number density of rare earth ions with increasing mole fraction of $Gd_2O_3$. Figure 11 shows the variations of polaron and ionic radius with changing mole fraction of rare earth oxide. Figure 12 shows the variations of polaron and ionic radius with changing number density of rare earth ions.

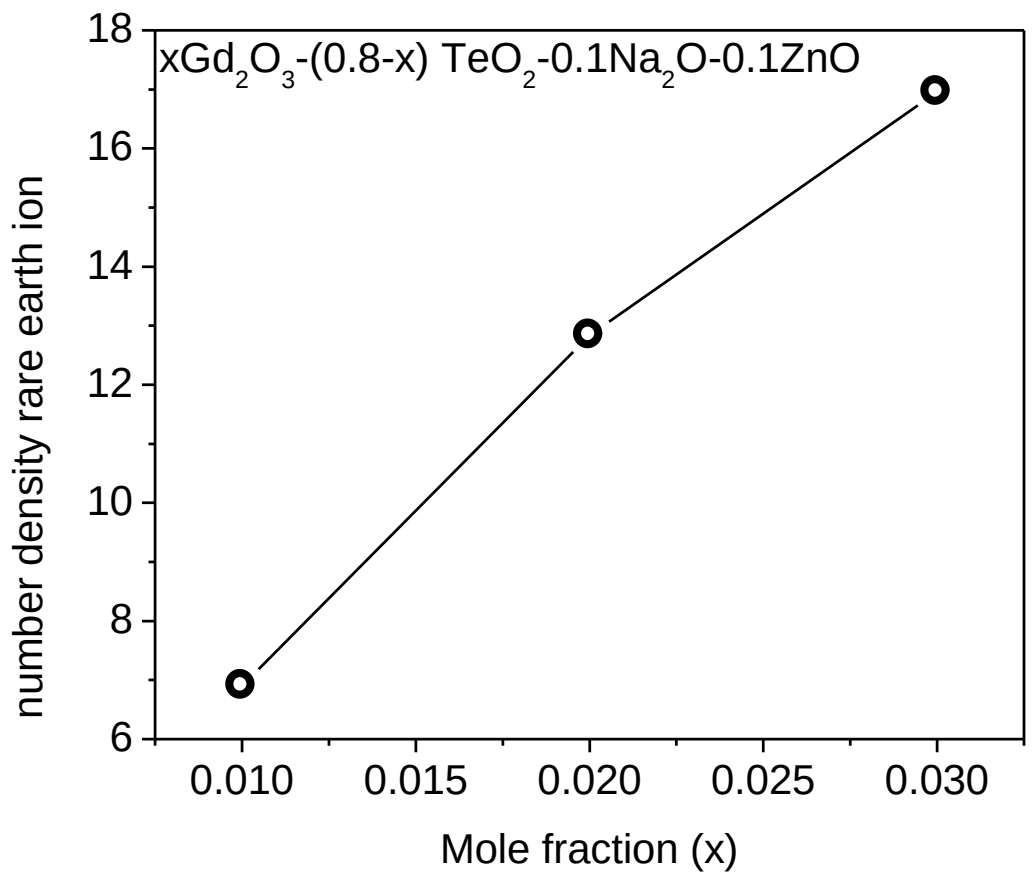

**Figure 10.** Influence of molar fraction of $Gd_2O_3$ on number density of rare earth ions of investigated glasses

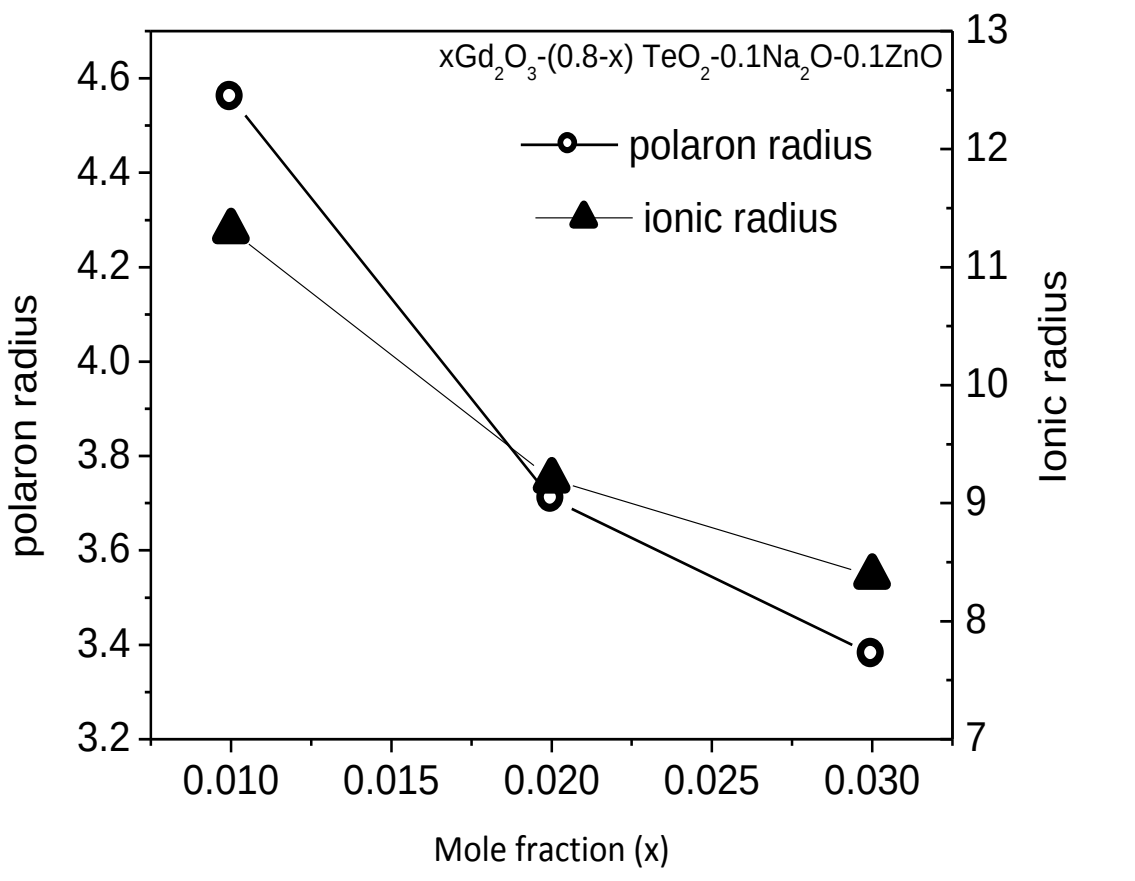

**Figure 11.** Influence of molar fraction of $Gd_2O_3$ onpolaron radii and ionic radii of rare earth ions of investigated glasses

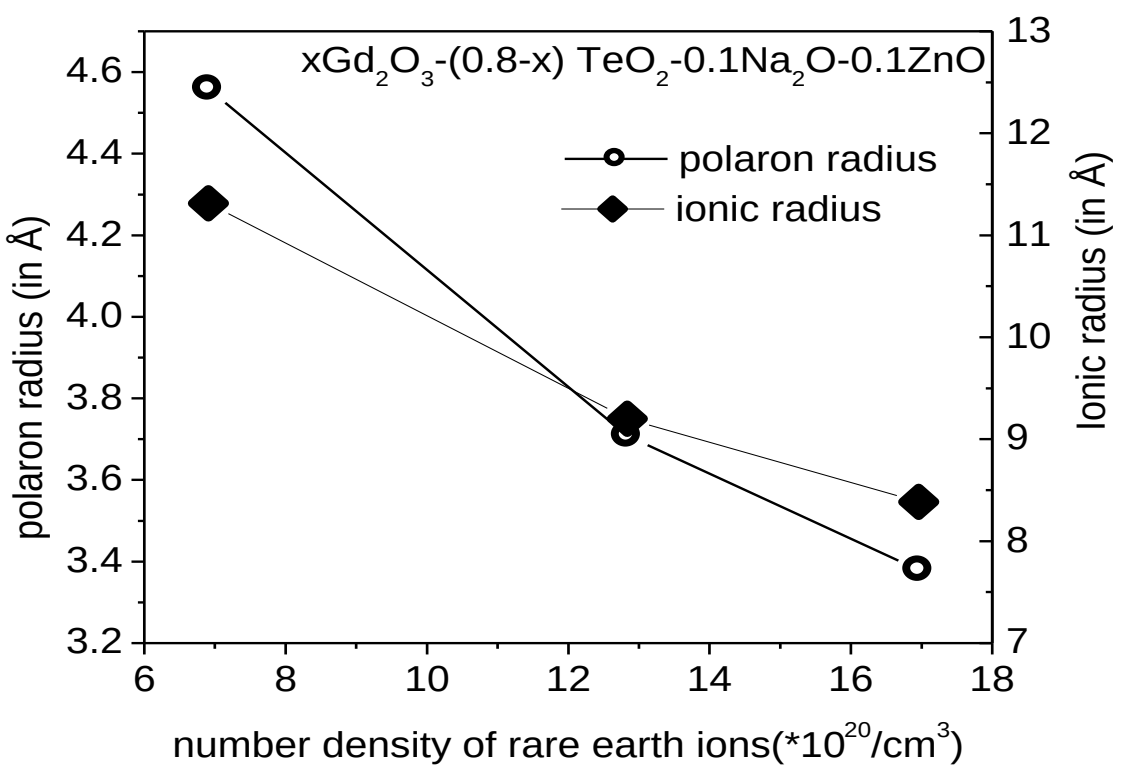

**Figure 12.** Influence of number density of rare earth ions on polaron radii and ionic radii of that ions for the investigated glasses

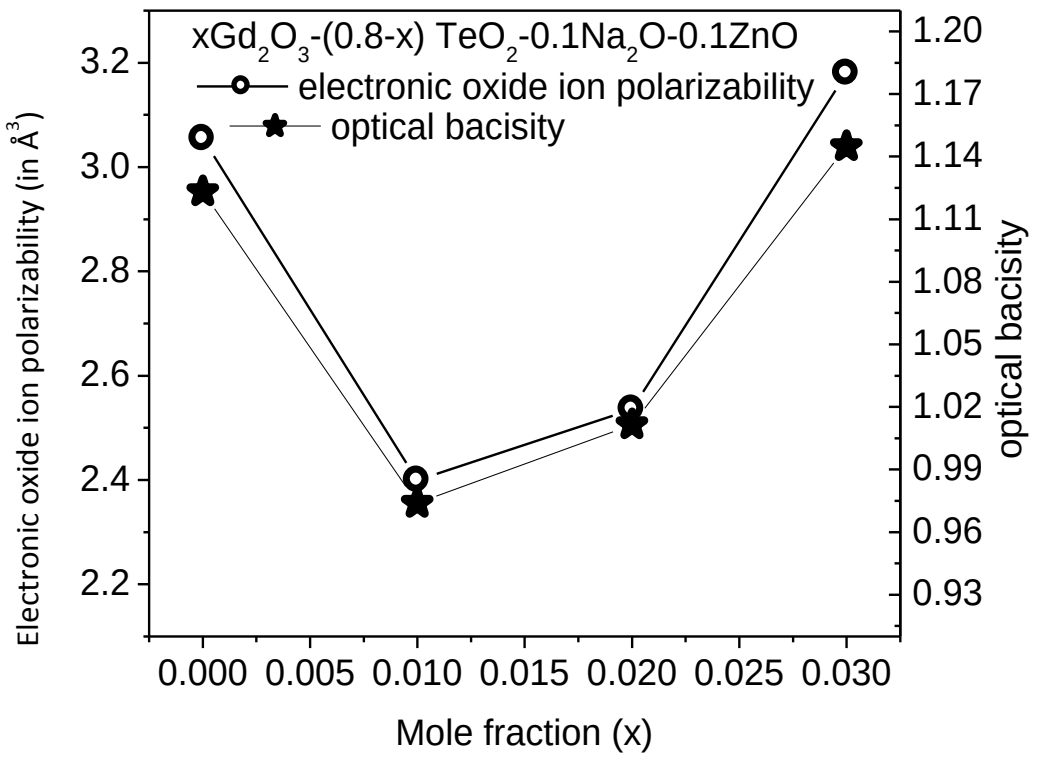

**Figure 13.** Influence molar fraction of $Gd_2O_3$ on electronic oxide ion polarizability and optical bacisity of the investigated glasses.

## Calculation of optical band gap based electronic oxide ion polarizability ($\alpha_{O2-}$) of oxide glasses and optical basicity (Λ).

### Electronic oxide ion polarizability ($\alpha_{O2-}$):

For a glass matrix X $A_pO_q$ - (1-X) $B_rO_s$ , oxide ion polarizability can be given by the relation [18, 25, 27], $\alpha_{O2-} = [(R_m/2.52-\sum\alpha_i](N_{o2-})^{-1}$ (23)
where $R_m= V_m\ [1-(E_g/20)^{1/2}]$ [28], $\sum\alpha_i= Xp\alpha_A+ (1-X)r\alpha_B$ = Molar cation polarizability. $\alpha_A$ , $\alpha_B$ are electronic polarizability of A and B cation respectively. $N_{02-}$= xq + (1-x)s =Number of oxide ions in chemical composition. Electronic polarizability values for different ions [$Na^+$ (0.181), $Zn^{2+}$ (0.286), $Te^{6+}$ (0.262), $Te^{4+}$ (1.595), $Gd^{3+}$ (0.61)] (all in unit $Å^3$) are taken from the paper reported by V. Dimitrov, T. Komatsu [27].

### Optical Basicity (Λ):

Duffy proposed a relation between optical basicity and electronic oxide ion polarizability as

$\Lambda=1.67*[1-(1/\ \alpha_{O2-})]$ (24)

Figure 13 shows the variations of electronic oxide ion polarizability and optical bacisity with changing molar fraction of $Gd_2O_3$.

## 3. Results & Discussions

### Table2

Measured values of density (ρ), Refractive index (n), molar volume ($V_m$), molar refraction ($R_M$), allowed indirect optical band gap ($E_g$), optical Dielectric constant ($\epsilon_r$), optical Susceptibility (χ), Metallization criterion (M’), polarizability ($\alpha_e$), Number density of rare-earth ions (N), Polaron radius ($r_p$), Inter ionic distance ($r_i$), Molar cation polarizability ($\sum\alpha_i$ ), Number of oxide ions in chemical composition ($N_{02-}$),optical band gap based electronic oxide ion polarizability ($\alpha_{O2-}$) and optical basicity (Λ) of glass samples.

| Sample | NZTG0 | NZTG1 | NZTG2 | NZTG3 |
|---|---|---|---|---|
| $\rho(gm\text{-}cm^{-3})$ | 4.596 | 5.506 | 5.189 | 4.634 |
| n | 2.188 | 2.149 | 2.201 | 2.209 |
| $V_m(cm^3)$ | 30.9 | 26.167 | 28.152 | 31.961 |
| $R_M(cm^3)$ | 17.242 | 14.313 | 15.821 | 18.026 |
| $E_g$(eV) | 3.907 | 4.108 | 3.842 | 3.804 |
| $\epsilon_r$ | 4.7875 | 4.6194 | 4.8447 | 4.8788 |
| X | 3.7875 | 3.6194 | 3.8447 | 3.8788 |
| (M’) | 0.441 | 0.453 | 0.438 | 0.436 |
| $\alpha_e(*10^{-24})\ cm^3$ | 6.834 | 5.673 | 6.271 | 7.145 |
| $N(*10^{20}/cm^3)$ | - | 6.9053 | 12.8367 | 16.9603 |
| $r_p$ (in Å) | - | 4.5594 | 3.7081 | 3.3793 |
| $r_i$ (in Å) | - | 11.3137 | 9.2013 | 8.3853 |
| $\sum\alpha_i$ (in $Å^3$) | 1.3444 | 1.33705 | 1.3369 | 1.33315 |
| $N_{02-}$ $(*10^{20}/cm^3)$ | 1.80 | 1.81 | 1.82 | 1.83 |
| $\alpha_{O2-}$(in $Å^3$) | 3.0543 | 2.3992 | 2.5359 | 3.1803 |
| Λ | 1.1232 | 0.9739 | 1.0115 | 1.1449 |

A series of $Gd_2O_3$ doped sodium zinc tellurite glass has been prepared by conventional melt-quenching technique. The amorphous nature of these glasses was confirmed by X-ray diffraction spectra taken at room temperature using Cu-$K_a$ radiation. Figure 2 shows the variation of Kubelka-Munk function with wavelength which is similar to the typical absorption spectra. Different optical parameters like band gap, refractive index, dielectric constant and susceptibility of these glasses are tabulated in Table 2. Figure 3 shows the calculation of optical band gap values of these glasses using the 'Kubelka-Munk' theory. Figure 4 and Figure 7 respectively shows the variation in the optical band gap and density of the glasses with increasing $Gd_2O_3$ contribution in the network. Such behaviour can be explained on the basis of decrease and increase in the non-bridging oxygen (NBO) content in the glass network [8]. In base glass (NZTGO) $Na_2O$ and ZnO are used as the network modifier.$Na^{2+}$ and $Zn^{2+}$ ions modify the tellurite glass structure by creating the NBO. In the beginning, for 0.01 mole fraction of $Gd_2O_3$, the non-bridging Oxygen (NBO) ion contents may decrease due to existance of gladolinium ion contents at interstitial sites of glass network making the structure more closed and hence, the sample becomes more insulating thereby increase in the_optical band gap and density has been observed. But in the later stage, the percentage of $Gd_2O_3$ is further increased (0.02 and 0.03 mole fraction) causing the increment of non bridging oxygen ion contents (NBO). Hence, due to more ionic character of NBO there is decrease in the optical band gap energies [12] and increase in NBO causes more open structure showing the decrease in density. Another reason behind the drop of band gap may be high dopant concentrations, which causes the broadening of the impurity band and the formation of band tails on the edges of the conduction and valence bands would lead to a reduction in $E_g$ as in semiconductors [23]. Figure 5 shows the variation of linear refractive index (n) with changing mole fraction of $Gd_2O_3$.

Table 2 shows the variation of metallization criterion with changing mole fraction of $Gd_2O_3$.In the beginning for 0.01 mole fraction width of conduction band decreases and hence metallization criterion has increased. Further increase in rare-earth content causes increment of conduction band width and decrement in metallization criterion on the basis of band gap. Decrease in metallization criterion indicates that the glass samples are metalizing [24].

We also show the variation of molar volume and polarizability respectively in Figure 7 and Figure 8 with increasing Gadolinium-oxide content in the glass composition. In the beginning molar volume was found to decrease for 1% $Gd_2O_3$ doping due to decrease in the bond length between two atoms compare to that of in base glass. Further increment of $Gd_2O_3$ (up to 3%) in the glass network shows the increment in the molar-volume due to increase in the bond-length between two atoms. Same trend is observed here for polarizability of oxide ions as non-bridging oxyzen (NBO) bonds have a much greater ionic character and lower bond energy profile; consequently NBO bonds have greater polarizability and cation refraction [8]. Figure 13 shows the variations of electronic oxide ion polarizability and optical bacisity with changing mole fraction of $Gd_2O_3$. The electronic oxide ion polarizability was found to decrease first for 1 % $Gd_2O_3$ doping and later increased with increase in $Gd_2O_3$ content in the glass network.The rise of electronic oxide ion polarizability with rare earth content may be due to the higher ability of oxide ions to transfer electrons to the surrounding cations.Same results were reflected for optical basicity.

## 4. Conclusions:

In these work we have found out the glass forming region for a series of $Gd_2O_3$ doped sodium zinc tellurite glasses. From the results above it can be seen that the optical band gapfor the above mentioned glasses can be tuned within the range from 3.804 to 4.108 eV depending on the $Gd_2O_3$ content of the glass samples. This suggests that these series of glasses are promising agents for UV-protective devices. Also the NBO density in these glasses seems to be controlled by the $Gd_2O_3$ content in them. Refractive indices of the prepared glasses are quite high as reported for tellurite glass [2]. In future one can study the changes in the relative population of the basic structural units for the tellurite matrix for different composition to find a correlation between the observed variations of the optical parameters that were obtained in the above study.